\documentclass[3p,times,twocolumn]{elsarticle}

\usepackage{ecrc}
\usepackage{epsfig}

\volume{00}

\firstpage{1}

\journalname{Nuclear Physics B Proceedings Supplement}

\runauth{P. Roig}

\jid{nppp}

\jnltitlelogo{Nuclear Physics B Proceedings Supplement}

\usepackage{amssymb}

\usepackage[figuresright]{rotating}

\begin{document}

\begin{frontmatter}



\dochead{}

\title{Lepton flavor violation in the Simplest Little Higgs model}


\author{P. Roig}

\address{Departamento de F\'isica, Centro de Investigaci\'on y de Estudios Avanzados del Instituto Polit\'ecnico Nacional,
Apartado Postal 14-740, 07000 Mexico City, Mexico}

\begin{abstract}
Little Higgs Models are a possible elegant solution to the hierarchy problem on the Higgs mass. As they predict naturally small deviations with respect to SM results, they are in agreement with all current experimental data. 
In this contribution, we review lepton flavor violation in the Simplest Little Higgs model focusing on semileptonic lepton flavor violating tau decays (where some new results are presented) and $H\to\ell\ell'$. Within this 
model, the most promising decay channels for discovering lepton flavor violation are $\mu\to e$ conversion in nuclei, $\mu\to e\gamma^{(*)}$, $\tau\to (e/\mu)\gamma$ and $\tau\to (e/\mu) (\pi^+\pi^-/\rho)$.
\end{abstract}


\end{frontmatter}


\section{Motivation}\label{Mot}
I will start recalling the case for searches and studies of lepton flavor violation and also the interest of analysing it within (the Simplest) Little Higgs models.

\subsection{Lepton Flavor Violation (LFV)}\label{LFV}
The discovery of neutrino oscillations evidences their non-vanishing mass and makes the charged lepton sector the only fermion subarea where flavor violation has not been unveiled yet. Moreover, the 
minimal extension of the SM with $3$ massive neutrinos predicts non-zero (although undetectable) branching ratios for charged LFV processes, i. e. $B(\mu\to e\gamma)\sim 10^{-54}$ \cite{mu2eg}. 
Therefore, the discovery of charged LFV would correspond, necessarily, to the effect of new dynamics.

For this reason there is an extensive hunt for new physics in searches for LFV muon, tau, Higgs, $Z^{(\prime)}$ decays and $\mu\to e$ conversion in nuclei; on which we had several experimental talks at 
this workshop \cite{exptalks} (see also the corresponding sections of \cite{Emilie}). Without dwelling in more detail into these, let us just highlight the impressive upper limit recently achieved by MEG 
$B(\mu\to e\gamma)\leq 4.2\cdot10^{-13}$ \cite{MEG}, which is a stringent constraint on new physics models. In parallel to this exhaustive experimental activity there is not a corresponding effort on the theory 
side (but for the easiest $L\to \ell \gamma^{(\star)}$ decays) and limited activity on the semileptonic LFV tau decays has been carried on \cite{TheoryLFV}.

LFV is not intrinsically related to any of the known problems of the Standard Model (SM): dark matter, baryon asymmetry of the universe, dark energy, (little) hierarchy problem, flavor problem, etc. However, 
it will be hopefully linked to any/some of them, so that its eventual measurement will shed light on any of these issues, helping to find the next standard theory.

\subsection{Simplest Little Higgs (SLH) model}\label{SLH}
Little Higgs models arise as an elegant solution to the (little) hierarchy problem on the Higgs mass: since the Higgs boson couples proportionally to others' particles masses, its mass would get huge quantum 
loop corrections in the presence of generic heavy new physics. Therefore, $m_H\sim 125$ GeV would need to result from an extreme fine-tuning among the diverse corrections. A theoretically beautiful solution 
to this problem is provided by Supersymmetry, but the absence of SUSY particles at a TeV questions that Nature chose this way. Another classical solution to the problem comes from the analog with QCD. Technicolor 
and its different evolutions again face naturalness problems when confronted to the lack of their imprints on LHC data. Still, the idea of composite Higgs models \cite{CompositeHiggs} can be the starting 
point to formulate a theory in which the Higgs boson is naturally light that accords with all present observations.

Scalar boson masses are not protected by any symmetry, however the pion is so light because it is the pseudo-Nambu-Goldstone boson of chiral symmetry breakdown. The idea of LH models is to justify the small 
Higgs mass similarly, as a consequence of the breaking of a global symmetry. These models assume a scale of compositeness $f$ (above which the new global symmetry is also displayed), which is much smaller than 
the electroweak vev ($f\geq 1$ TeV) and the structure of the model is arranged so that the Higgs mass is radiatively generated. There are new 'little' particles with masses of $\mathcal{O}(f)$ and the UV completion 
of the model is expected at some tens of TeVs, where the theory would become strongly coupled ($4\pi f\gtrsim 12$ TeV). Thus we can expand perturbatively our amplitudes in $v/f$ and keep only the leading term.

Among the LH models there are product group $(\left[SU(2)\otimes U(1)\right]^N)$ and simple group models ($SU(N)\otimes U(1)$). Since the former need and ad-hoc symmetry (T-parity) to solve the hierarchy 
problem, we will take the simplest of the latter ($N=3$) for our study of LFV tau \cite{Lami:2016vrs} and Higgs decays \cite{OurLetter} that we present in Sects. \ref{LFVtau} and \ref{LFVHiggs} preceded by 
a short account on the SLH model next (see also \cite{PosterSLH}).

\section{A brief sketch of the SLH model}\label{Sketch}
The symmetry structure of the SLH model \cite{SLH_bib} is given by $\left[SU(3)\otimes U(1)\right]_1\otimes \left[SU(3)\otimes U(1)\right]_2$, where only the diagonal group is gauged. There are two different 
symmetry breakdowns (requiring two complex scalar fields, triplets under $SU(3)_1$ and $SU(3)_2$, respectively): on the one hand the gauged diagonal subgroup is broken down to the SM electroweak gauge group, 
yielding 5 Goldstone bosons which give mass to the additional 'little' gauge bosons (among these only $W^{\prime\pm}$ and $Z^\prime$ play a r\^ole in our study). On the other hand, the global symmetry is 
broken similarly, with associated Goldstone bosons including the Higgs degrees of freedom.

Every fermion family contains a left-handed triplet (adding to the SM doublets one 'little' particle) and corresponding singlets. Heavy neutrinos, $N_k$ ($k=1,2,3$) are fundamental to our study, since 
they drive the LFV through their couplings. Though the quark sector is not unambiguously defined, we will follow the anomaly-free embedding for the new quarks. Under reasonable assumptions, only the 
first generation 'little' quark, $D$, matters to our discussion \cite{delAguila:2011wk}.

\section{LFV tau decays}\label{LFVtau}
We presented our results for $\tau\to \ell (P/PP/V)$ ($\ell=e,\,\mu$; $P$ is short for pseudoscalar meson and $V$ for vector resonance) in Ref.~\cite{Lami:2016vrs}~\footnote{The SLH preserves lepton universality. As a result,
we obtain the same branching ratios irrespective of $\ell=e,\,\mu$.}. There we decided to include the effect 
of only two heavy neutrinos in our analysis. This corresponds to the case where there is a GIM-like mechanism acting in the mixing matrix among charged leptons and heavy neutrinos which effectively 
decouples $N_3$. In this scenario, also the contributions of $N_1$ and $N_2$ cancel each other partially (according to the similarity of their masses). Here, instead, we will consider the most general 
scenario where no particular pattern of this mixing matrix is assumed. Generally, this will increase our predicted LFV observables.

\begin{figure}[h!]
\centerline{\includegraphics[width=\linewidth]{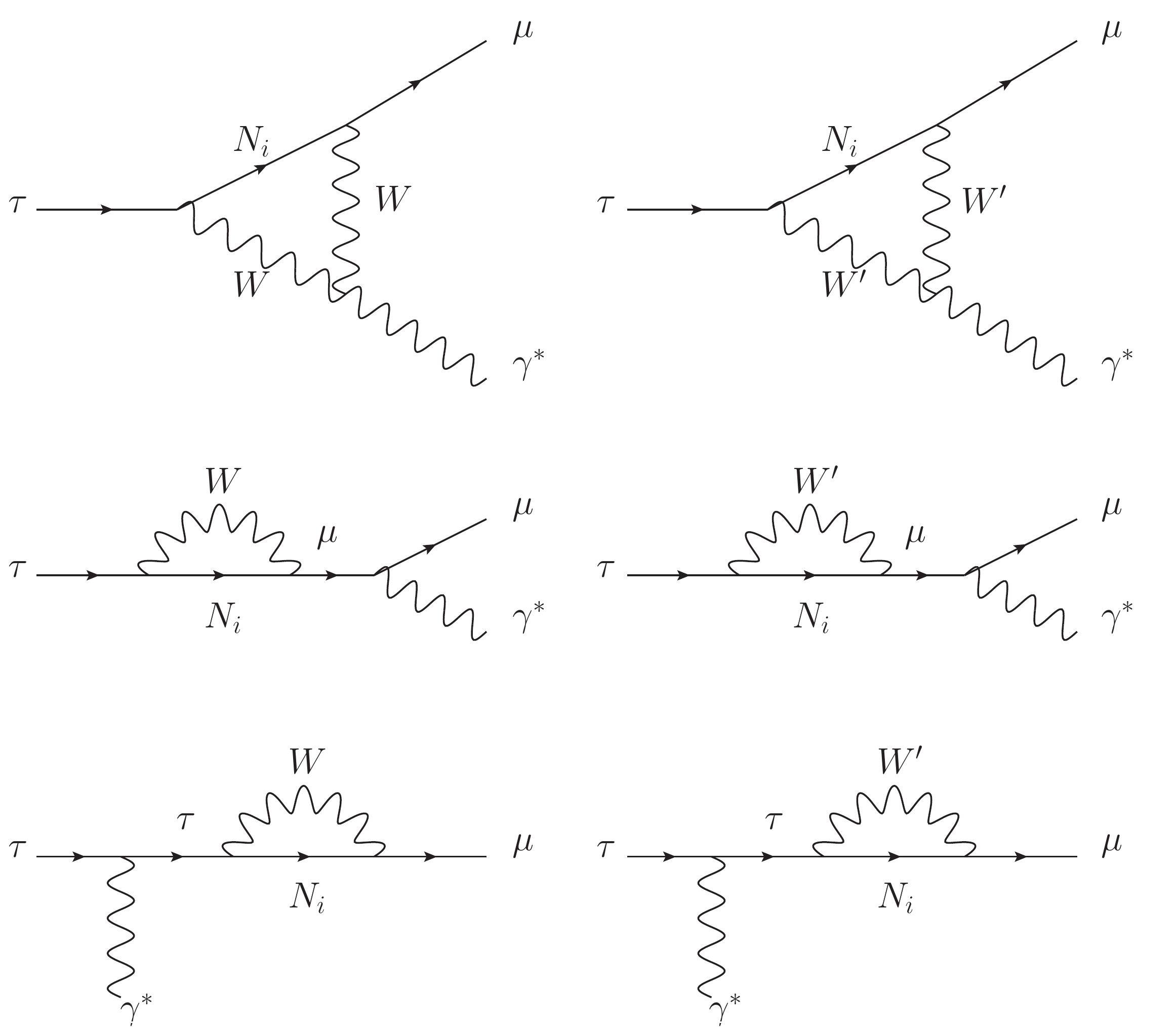}}
\caption{\label{figgamma} Penguin-like diagrams for $\tau \rightarrow \mu \gamma^*$ in the SLH model.}
\end{figure}

\begin{figure}[h!]
\centerline{\includegraphics[width=\linewidth]{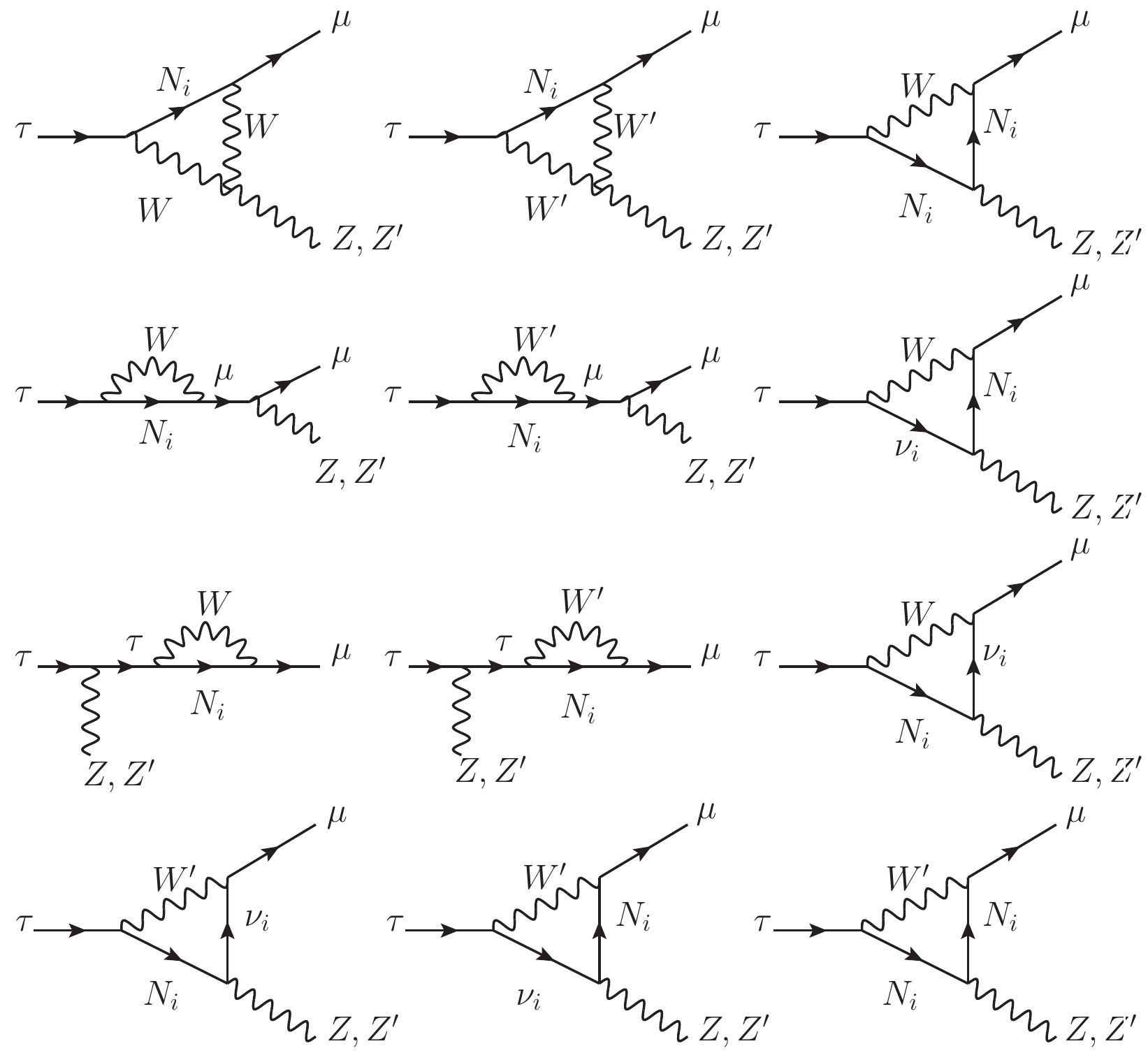}}
\caption{\label{figZ} Penguin-like diagrams for $\tau \rightarrow \mu (Z/Z^\prime)$ in the SLH model.}
\end{figure}

\begin{figure}[h!]
\centerline{\includegraphics[width=0.8\linewidth]{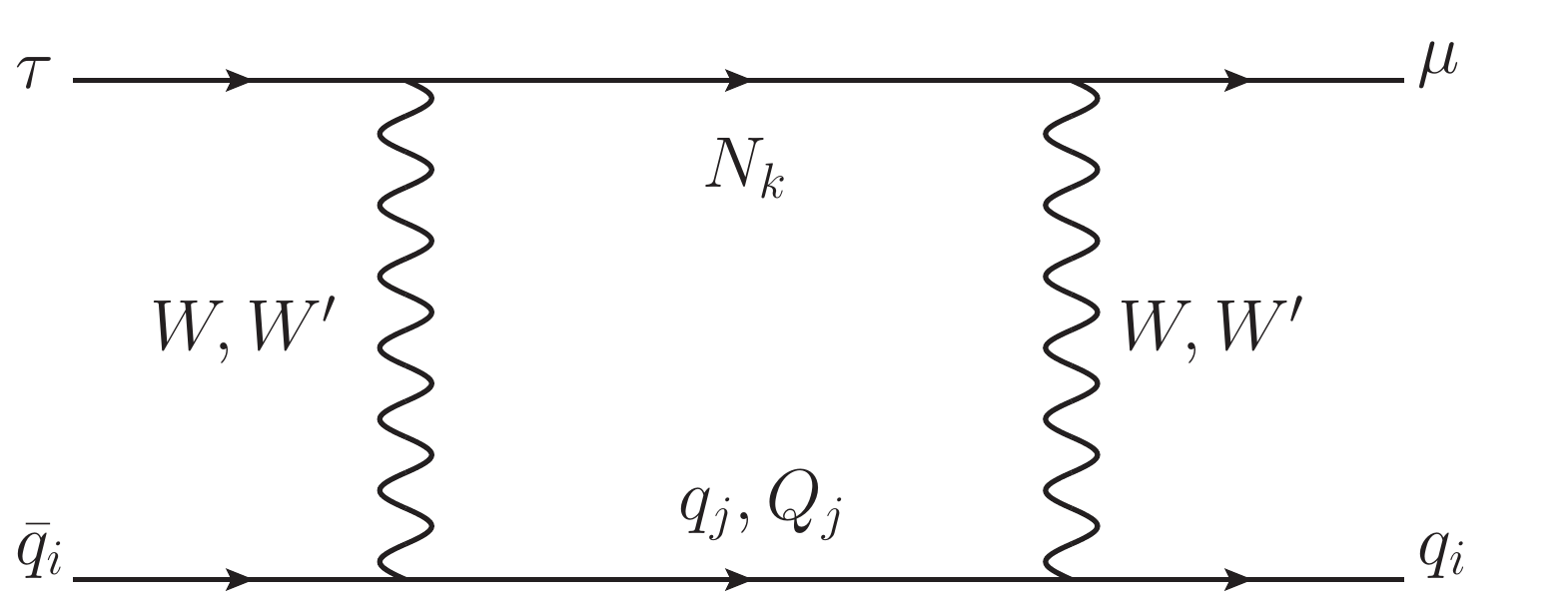}}
\caption{\label{figbox} Box diagrams for $\tau \rightarrow \mu q \overline{q}$ in the SLH model. The internal quark states are $(u,\overline{u}) \rightarrow 
\{d, D\}$, 
$(d, \overline{d}) \rightarrow \{u \}$, $(s, \overline{s}) \rightarrow \{c\}$.} 
\end{figure}

The one-loop diagrams contributing to these decays can be seen in figures \ref{figgamma}, \ref{figZ} and \ref{figbox} (with the $\gamma-$, $Z^{(\prime)}-$ and box-mediated contributions, respectively). We 
have computed them in the unitary gauge, where only physical particles appear. As a result, the cancellation of divergences becomes subtle, and the sum of the divergences of the penguin-like diagrams is 
cancelled by that of the box contributions. Parity forbids $\gamma$-mediated contributions to the processes with one pseudoscalar meson. However, since these and the box-mediated contributions turn out to 
be of similar magnitude \footnote{$Z$ and $Z^\prime$ contributions are negligible in all cases.} $\tau\to \ell P$ decays are predicted at a comparable rate to the $\tau\to \ell (PP/V)$ processes. Along our 
computation we have kept the leading term in the expansion parameter $v/f$ and set $m_\nu$, $m_\ell$ and $M_\tau$ to zero \footnote{$M_\tau$ also sets the largest scale of external momenta, which are then 
negligible in the evaluation of the loop integrals.}.

As a result of the embedding of the SM group into the SLH group, the only couplings entering the amplitudes are the SM $SU(2)$ and $U(1)$ couplings. In addition, the expressions depend on the ratios 
$\chi_j\equiv M_{N_j}^2/M_{W^\prime}^2\sim\mathcal{O}(1)$ and $\omega\equiv M_W^2/M_{W^\prime}^2<<1$. $\delta_\nu=-v/(\sqrt{2}f\rm{tan}\beta)$ appears in the change between the flavor and mass bases for 
the (light and heavy) neutrinos and turns out to be an important parameter allowing to set the bound $f\rm{tan}\beta\leq 3.48$ TeV \cite{delAguila:2011wk} (in the two-heavy neutrino scenario) studying 
$\mu\to e\gamma$, $\mu\to eee$ and $\mu N\to e N$ in the SLH model~\footnote{A thorough discussion of the phenomenological relevance of these decays modes can be found in this reference.}.

The remaining expressions contain quark bilinear currents which still need to be hadronized to make contact with the experimental searches. This is done in an essentially model-independent way, writing 
those fermion bilinears in terms of the QCD quark currents and proceeding to their hadronization guided by chiral symmetry \cite{ChPT}, axiomatic field theory properties implemented naturally through 
dispersion relations \cite{Sergi} and the QCD asymptotics \cite{BrodskyLepage}, benefitting as well from the precise data at our disposal on two-meson factors. For the these, we use the expressions 
given in Refs.~\cite{2mesonFFs} \footnote{In the one-meson case, the hadronization is encoded in the corresponding meson decay constant and the $\eta-\eta^\prime$ mixing parameters \cite{Lami:2016vrs}. 
The $V$ cases can be obtained by considering the dominant $PP$ channel with the di-meson system around the mass and width of the vector resonance ($\pi^+\pi^-$ for the $\rho$ and $K\bar{K}$ for the 
$\phi$).}.

\begin{figure}[h!]
\centerline{\includegraphics[width=\linewidth]{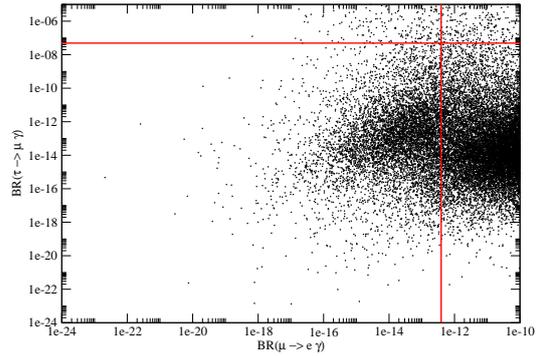}}
\caption{\label{figtaumu} $B(\mu \rightarrow e \gamma)$ vs. $B(\tau \rightarrow \mu \gamma)$ in the SLH model. The solid lines signal the upper limit at 90$\%$ C. L. for each mode.}
\end{figure}

For our phenomenological analysis within the SLH model, we have varied randomly the model parameters in the ranges $2$ TeV $\leq\, f\leq 10$ TeV, $1\leq$tan$\beta\leq 10$, keeping its product below $3.5$ TeV. 
In the mixing matrix between charged leptons and heavy neutrinos we have neglected CP violation but kept it general otherwise. We have allowed for a factor of up to ten in the ratio between successive 
heavy neutrino masses and verified that all LFV low-energy constraints were satisfied before admitting a point in the model's parameter space. This restriction is needed, as can be seen from 
fig.~\ref{figtaumu}.

\begin{figure}[h!]
\centerline{\includegraphics[width=\linewidth]{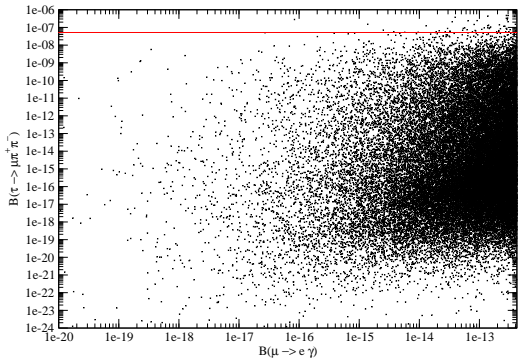}}
\caption{\label{fig2a} Correlation between $B(\mu \rightarrow e \gamma)$ and $B(\tau \rightarrow \mu \pi^+ \pi^-)$ in the SLH model.}
\end{figure}

\begin{figure}[h!]
\centerline{\includegraphics[width=\linewidth]{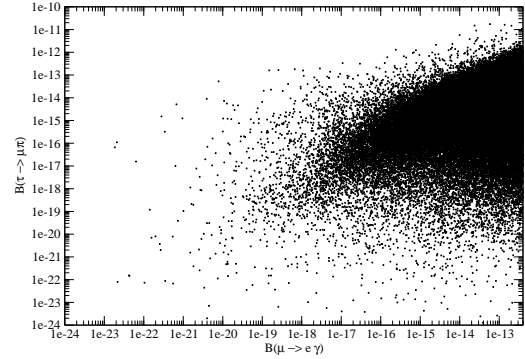}}
\caption{\label{fig2b} Correlation between $B(\mu \rightarrow e \gamma)$ and $B(\tau \rightarrow \mu \pi^0)$ in the SLH model.}
\end{figure}

Within the SLH model, the correlation between the most restricting low-energy process and the most abundant one- and two-meson LFV tau decays \footnote{$B(\tau \rightarrow \mu \pi^+ \pi^-)$ and $B(\tau \rightarrow \mu \pi^0)$ are extremely 
correlated, as can be seen in \cite{Lami:2016vrs}. This is a result of the hadronization process, as the $\pi^+\pi^-$ channel is driven by the $\gamma$-exchange, while the one-pion mode is saturated by the box contribution.} is plotted in figures 
\ref{fig2a} and \ref{fig2b}, respectively. In both figures the x-axis is cut at the 90$\%$ C.L. upper limit for $B(\mu\to e \gamma)$. The solid line in Fig. \ref{fig2a} indicates the corresponding upper limit for $B(\tau \rightarrow \ell \pi^+ \pi^-)$.  
A similar line is not shown on figure \ref{fig2b} because if the other low-energy restrictions on LFV processes are fulfilled, $B(\tau \rightarrow \ell \pi^0)$ is at least four orders of magnitude below 
its corresponding upper bound. Thus, in the SLH model, the only semileptonic LFV tau decays that can compete with $\mu N\to e N$, $\mu\to e\gamma$ and $\tau\to\ell\gamma$ as golden channels for the detection of LFV are 
$\tau \rightarrow \ell \pi^+ \pi^-$ and $\tau \rightarrow \ell \rho$ (which is only a factor $\sim 2$ smaller than the $\pi^+\pi^-$ mode).

Three-dimensional plots that allow to represent the simultaneous dependence of the branching ratios on two model parameters can be found in my talk's file \cite{MyTalk}. This, however, does not yield any new information (provided 
the GIM-like suppression is understood) with respect to the most conventional 2-D plots. Then, the dependence of the results on the model parameters is basically the one found for the case with only two effective heavy 
neutrinos in the 2-D plots of Ref.~\cite{Lami:2016vrs}: results depend quite mildly on $f$, tan$\beta$, max$\left\lbrace|V^{i\mu}V^{i\tau}|\right\rbrace$ (sin$2\theta$ for the GIM-like scenario) and the heavy neutrino 
spectroscopy. As expected, BRs:\\
- decrease with f according to the dependence of the amplitude on $(v/f)^2$,\\
- are almost constant for tan$\beta\geq3$, while they exhibit a marked narrow dip around tan$\beta=2$, where the BR is reduced by an order of magnitude.\\
- increase as sin$^22\theta$ (similarly for max$\left\lbrace|V^{i\mu}V^{i\tau}|^2\right\rbrace$).\\
- vary smoothly with the neutrino masses hierarchy. In the GIM-like case, the suppression of the BRs gets stronger for $M_{N_1}\sim M_{N_2}$.

\section{LFV Higgs decays}\label{LFVHiggs}
\begin{figure}[h!]
\centerline{\includegraphics[width=\linewidth]{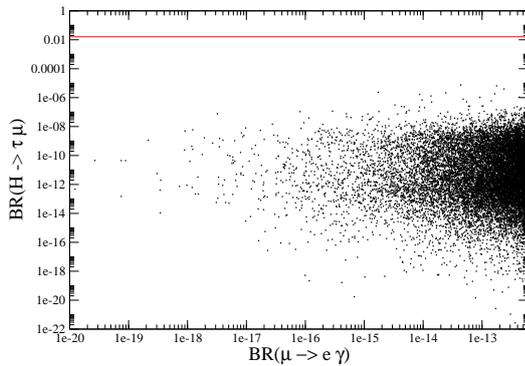}}
\caption{\label{figHiggs} Correlation between $B(\mu \rightarrow e \gamma)$ and $B(H \rightarrow \tau \mu)$ in the SLH model. The solid line shows the 95$\%$ C.L. upper limit set by the LHC experiments.}
\end{figure}

It has not been necessary to update our analyses of $H\to\ell\ell'$ \cite{OurLetter}. The dependence on the model parameters follows the patters explained in Sect.~\ref{LFVtau}. As it can be seen in fig.~\ref{figHiggs}, even in the case with three 
active heavy neutrinos (where the considered LFV Higgs decays BRs are four orders of magnitude larger than in the GIM-like scenario \cite{OurLetter}) the SLH predicts unmeasurable BRs at LHC, provided the low-energy 
constraints on LFV processes are satisfied. Similar small BRs for these decays have been found recently within LH models \cite{LHresults}~\footnote{SLH models also predict generally small departures from the SM in Higgs 
couplings, which are in good agreement with present measurements \cite{Han:2013ic}.}. We refer the reader to Ref.~\cite{OurLetter} for a complete discussion of our results on LFV Higgs decays.

\section{Conclusions}\label{Concl}
Little Higgs models (particularly SLH) remain as elegant candidates to alleviate the hierarchy problem on the Higgs mass, respecting all experimental bounds. (S)LH models predict small LFV decay rates which could escape detection 
at Belle-II and (specially) at LHC. Within SLH, LFV detection should be easier for a general (not GIM-like) pattern of the 3 heavy neutrinos of the model. In that case, the most promising channels for its discovery would be $\mu N\to e N$, 
$\mu\to e\gamma$, $\tau\to (e/\mu)\gamma$ and $\tau\to (e/\mu) (\pi^+\pi^-/\rho)$.

\section*{Acknowledgements}The author wants to thank and congratulate the local and international advisory Committees for their organization of the fruitful TAU'16 Workshop. I acknowledge the collaboration 
with Jorge Portol\'es and Andrea Lami in the research reported in this proceedings contribution. I am grateful to Jorge Portol\'es for his critical reading of the manuscript of this text. 




\begin{thebibliography}{00}
\bibitem{Sergi}
  Talks given by Sergi Gonz\`alez-Sol\'is (one replacing Rafel Escribano, who unfortunately could not finally attend the conference) at the TAU'16 Workshop, these proceedings.
\end{thebibliography}



\end{document}